\documentclass[sigconf]{acmart}

\AtBeginDocument{%
  \providecommand\BibTeX{{%
    \normalfont B\kern-0.5em{\scshape i\kern-0.25em b}\kern-0.8em\TeX}}}
    
\acmConference[IUI 2020]{ACM International Conference on Intelligent User Interfaces}{March 17--20, 2020}{Cagliari, Italy}
\acmBooktitle{IUI '20: ACM International Conference on Intelligent User Interfaces,
  March 17--20, 2020, Cagliari, Italy}

\copyrightyear{2020}
\acmYear{2020}
\setcopyright{acmcopyright}
\acmConference[IUI '20]{25th International Conference on Intelligent User Interfaces}{March 17--20, 2020}{Cagliari, Italy}
\acmBooktitle{25th International Conference on Intelligent User Interfaces (IUI '20), March 17--20, 2020, Cagliari, Italy}
\acmPrice{15.00}
\acmDOI{10.1145/3377325.3377512}
\acmISBN{978-1-4503-7118-6/20/03}

\usepackage{balance}       
\usepackage{graphics}      
\usepackage{mathptmx}
\usepackage{booktabs}
\usepackage{textcomp}
\usepackage{algorithm}
\usepackage{algorithmic}
\usepackage{amsmath}
\usepackage{caption}
\usepackage{subcaption}
\usepackage{soul}
\usepackage{xcolor}

\renewcommand{\textrightarrow}{$\rightarrow$}

\def\authornote#1#2#3{{\color{#2}{\textsl{\small#1:[*#3*]}}}}

\newenvironment{flushitemize}{%
\begin{list}{$\bullet$}
   {\setlength{\leftmargin}{15pt}}%
    \setlength{\labelwidth}{20pt}
    \setlength{\itemindent}{0pt}
    \setlength{\labelsep}{0.5em}
 \setlength{\itemsep}{1pt}
 \setlength{\parskip}{0pt}
 \setlength{\parsep}{0pt}}
 {\end{list}}
\begin{document}
\raggedbottom
\title{Leveraging Rationales to Improve Human Task Performance}


\author{Devleena Das}
\affiliation{%
 \institution{Georgia Institute of Technology}
 \country{Atlanta, Georgia}}
 \email{ddas41@gatech.edu}
\author{Sonia Chernova}
\affiliation{%
 \institution{Georgia Institute of Technology}
 \country{Atlanta, Georgia}}
 \email{chernova@gatech.edu}


\begin{abstract}
Machine learning (ML) systems across many application areas are increasingly demonstrating performance that is beyond that of humans.  In response to the proliferation of such models, the field of Explainable AI (XAI) has sought to develop techniques that enhance the transparency and interpretability of machine learning methods. In this work, we consider a question not previously explored within the XAI and ML communities: \textit{Given a computational system whose performance exceeds that of its human user, can explainable AI capabilities be leveraged to improve the performance of the human?} We study this question in the context of the game of Chess, for which computational game engines that surpass the performance of the average player are widely available.  We introduce the \textit{Rationale-Generating Algorithm}, an automated technique for generating rationales for utility-based computational methods, which we evaluate with a multi-day user study against two baselines. The results show that our approach produces rationales that lead to statistically significant improvement in human task performance, demonstrating that rationales automatically generated from an AI's internal task model can be used not only to \textit{explain} what the system is doing, but also to \textit{instruct} the user and ultimately improve their task performance.
\end{abstract}




\maketitle

\section{Introduction}

Machine learning (ML) systems across many application areas are increasingly demonstrating performance that is beyond that of humans.  From games, such as Chess \cite{hsu1999ibm} and Atari \cite{montfort2009racing}, to high-risk applications, such as autonomous driving \cite{teichmann2018multinet} and medical diagnosis \cite{kermany2018identifying}, human users are becoming increasingly surpassed by, and reliant on, autonomous systems.  In response to the proliferation of such models, the field of Explainable AI (XAI) has sought to develop techniques that enhance the transparency and interpretability of machine learning methods.  XAI approaches have included taking advantage of intrinsically interpretable models (e.g., decision sets) \cite{lakkaraju2016interpretable, letham2015interpretable}, as well as developing interpretable approximations to explain the behavior of non-interpretable black-box models (e.g., decision tree approximations of deep neural networks) \cite{ribeiro2016should,10.5555/3305381.3305576}.
The vast majority of XAI research has focused on expert users, for example, medical personnel evaluating the decision-making capability of an automated diagnosis system \cite{adadi2018peeking,ribeiro2016should}.  

In this work, we consider a question not previously explored within the XAI and ML communities: \textbf{Given a computational system whose performance exceeds that of its human user, can explainable AI capabilities be leveraged to \textit{improve} the performance of the human?} In other words, can the increasingly powerful machine learning systems that we develop be used to, in turn, further human capabilities?  Within the context of XAI, we seek not to explain to the user the inner workings of some algorithm, but to instead communicate the rationale behind a given decision or choice.  The distinction between \textit{explanation} and \textit{rationale} has been previously introduced by Ehsan et al. \cite{ehsan2019automated}, who defined explanations as a way to expose the inner workings of a computational model through any communication modality (e.g., visual heatmap \cite{fong2017interpretable}), often in a way that is accessible only to experts.  Rationales, on the other hand, are defined as natural language explanations that do not literally expose the inner workings of an intelligent system, but instead provide contextually appropriate natural language reasons. These natural language reasons are accessible and intuitive to non-experts (e.g., ``I had to go forward to avoid the red vehicle.''),
facilitating understanding and communicative effectiveness. Within the context of their work, Ehsan et al. introduced a computational method for automatically generating rationales and validated a set of human factors that influence human perception and preferences (i.e., contextual accuracy, intelligibility, awareness, reliability, strategic detail) ~\cite{ehsan2019automated}.

In this work, we explore whether providing human users with a \textit{rationale} of an intelligent system's behavior can lead to improvement in the user's performance.  We study this question in the context of the game of Chess, for which computational game engines that surpass the performance of the average player are widely available.  Our work makes the following contributions: 
\begin{flushitemize}
\item We introduce the \textit{Rationale-Generating Algorithm (RGA)}, an automated technique for generating rationales for utility-based computational methods. 
\item We study two variants of our approach, one that takes advantage only of the system's knowledge (RGA), and a second (RGA+) that additionally incorporates human domain expert knowledge. 
\end{flushitemize}
We evaluate both techniques in a multi-day user study, comparing against two baselines to measure human learning performance. The results demonstrate that our approach produces rationales that lead to improvement in human task performance in the context of chess endgames. Using winning percentage and percentile rank of player moves as a measure of performance, we observe that the inclusion of domain expert knowledge (RGA+) significantly improves human task performance over both baselines when compared to non-rationale baselines. Additionally, user self-reported performance ratings also show that rationale-based interfaces lead to greater perceived user understanding of the task domains than non-rationale baselines.  In summary, our approach is the first to demonstrate that rationales automatically generated from an AI's internal task model can be used not only to \textit{explain} what the system is doing, but also to \textit{instruct} users in a manner that ultimately improves their own task performance in the absence of rationales.

\vspace{-.1cm}

\section{Related Works}
Traditionally, in the XAI community, an interpretable model can be described as one from which an AI expert or user can deduce model performance based on given inputs \cite{ehsan2019automated}. The methods of interpretability can vary based on the complexity of a model and a wide range of survey papers summarize the different XAI models currently developed \cite{adadi2018peeking,zhang2018visual,ribeiro2016should}. While some of the existing models are inherently interpretable, suitable for model-intrinsic decision-making \cite{letham2015interpretable,caruana2015intelligible}, other complex models need model-agnostic approaches for interpretability \cite{10.1145/3236009,zhang2018interpreting, zhang2019interpreting, wu2018beyond}. Despite the differing approaches for interpretability, XAI models have the common motivation of improving human understanding of AI systems and building trust with these systems \cite{gunning2019darpa,gunning2017explainable}. 

One existing method for developing interpretability is to use intrinsic models in which interpretability is embedded inside the model. In \cite{letham2015interpretable} a Bayesian Rule List (BRL) is created to produce posterior distributions over permutations of \textit{'if, then, else'} rules for a single prediction classification. Since these decision lists are intrinsically understandable, using them as the basis of BRL is successful in developing interpretability. However, the accuracy of BRL depends highly on the Bayesian prior favoring concise decision lists and a small number of total rules, limiting the applicability of this approach. To extend the applicability of model-intrinsic methods beyond the scope of concise decision lists, a generative additive model (GAM) is created to provide both high accuracy (better than random forest, logitboost and SVMs) and high interpretability for a single prediction classification \cite{caruana2015intelligible}. To achieve model understanding, GAM creates a visual representation of the pairwise factors that result in a prediction and provides an ability for modular testing. \cite{caruana2015intelligible} refers to modular testing as allowing model-experts to easily remove and insert individual factors or pairwise factors to examine their effects on a prediction. While methods such as GAM provide interpretability for regression and some classification models, the intrinsic nature of its model makes it hard to provide interpretability for more black-box models which need model-agnostic implementations of interpretability.

An alternative to the model-intrinsic approach described above is a model-agnostic method for interpretability. Model-agnostic methods are
a significant focus within the XAI community since model-agnostic methods provide high performance accuracy, but do not allow for inherent decoding and visualization of model prediction. Most model-agnostic methods produce interpretability by developing surrogate models, post-hoc implementations of interpretability derived from inherently interpretable models. In \cite{10.1145/3236009} an ad-hoc genetic algorithm is used to generate neighbor nodes for a specific local instance, which is then trained by a decision tree classifier to generate a logic rule. This logic rule represents the path in the decision tree that explains the factors that lead to a prediction. Additionally, \cite{zhang2019interpreting} uses inherently explainable decision tree models to learn CNN layers in order to provide CNN rationales at the semantic level. \cite{zhang2019interpreting} defines a CNN rationale as an ability to quantify the set of objects that contribute to a CNN prediction from prediction scores. \cite{zhang2018interpreting} also uses the concepts of explanatory graphs to visually reveal the hierarchy of knowledge inside CNNs in which each node summarizes the knowledge in the feature maps of a corresponding conv-layer. However, the decision tree rationales and explanatory graphs only provide quantitative distributions and visualizations to domain-experts, and are not interpretable to general users without domain knowledge. Moreover, the complexity of these decision trees vary greatly based on the complexity of the model, sometimes no longer staying interpretable due to a large node space. Thus, \cite{wu2018beyond} creates a tree regularization penalty function that helps produce interpretability of complex models by using moderately-sized decision trees. Altogether though, the focus of the intrinsic and agnostic methodologies remain largely focused on giving insight into the inner workings of the AI systems, rather than providing humanly understandable rationales that extend beyond domain-expert usability.


The HCI community acknowledges the gap between XAI and the human usability of XAI, communicating that the explanations from AI and ML communities do not yet possess a large-scale efficacy on human users \cite{abdul2018trends}. To address the importance of user-friendly explanations for any user, not only domain-experts, researchers in HCI have developed context-aware rules to focus interest on easy-to-use interfaces that are aware of their environment and context of use \cite{dey2018context}. However, many of these existing context-aware rules are developed as frameworks to aid decision-making in domain-specific applications such as smart homes \cite{costanza2014doing, bourgeois2014conversations} and the office \cite{cheverst2005exploring}, instead of utilizing a model-agnostic approach encouraged by the AI/ML community. In addition to the context-aware rules used above, educators have focused on intelligent tutoring systems (ITS) \cite{mahdi2016intelligent, boyer2011investigating, hilles2017knowledge} to generate explanations for learning. While we similarly seek to improve the performance of a non-expert user, our work differs from tutoring systems in that the rationales produced by our system are generated based on and reflect an AI's internal computational model, not a prescribed curriculum.

In the context of exploring effective methods of representing AI model predictions to human users, \cite{feng2019can} investigates the helpfulness of different factors that can help interpret an AI decision. The author focuses on methods such as presenting evidence examples, highlighting important input features and visualizing uncertainty in the context of the trivia game Quizbowl. Human performance improvement is analyzed by the time taken to respond to a question and the accuracy of an answer. However, since improvement is always measured in the presence of interpretations, their work does not provide insight on whether long-term human knowledge and task performance increase in the absence of the assistance provided by the interface.


To address a more generic solution for generating human understandable explanations from AI/ML produced models outside of a specified curriculum, \cite{ehsan2019automated} develops human understandable rationale in the context of the game Frogger. The rationale generation uses natural language processing to contextualize state and action representations for a reinforcement learning system and studies the human factors that influence explanation preferences. The metrics for measuring interpretability of the rationales are primarily focused on perceived understanding of the rationale. In contrast, we generate rationales and measure task performance and self-perceived task performance improvement, as opposed to surveying understanding of a given rationale. 


\section{Rationale Generating Algorithm}

\begin{figure}
\centering
  \includegraphics[width=0.9\columnwidth]{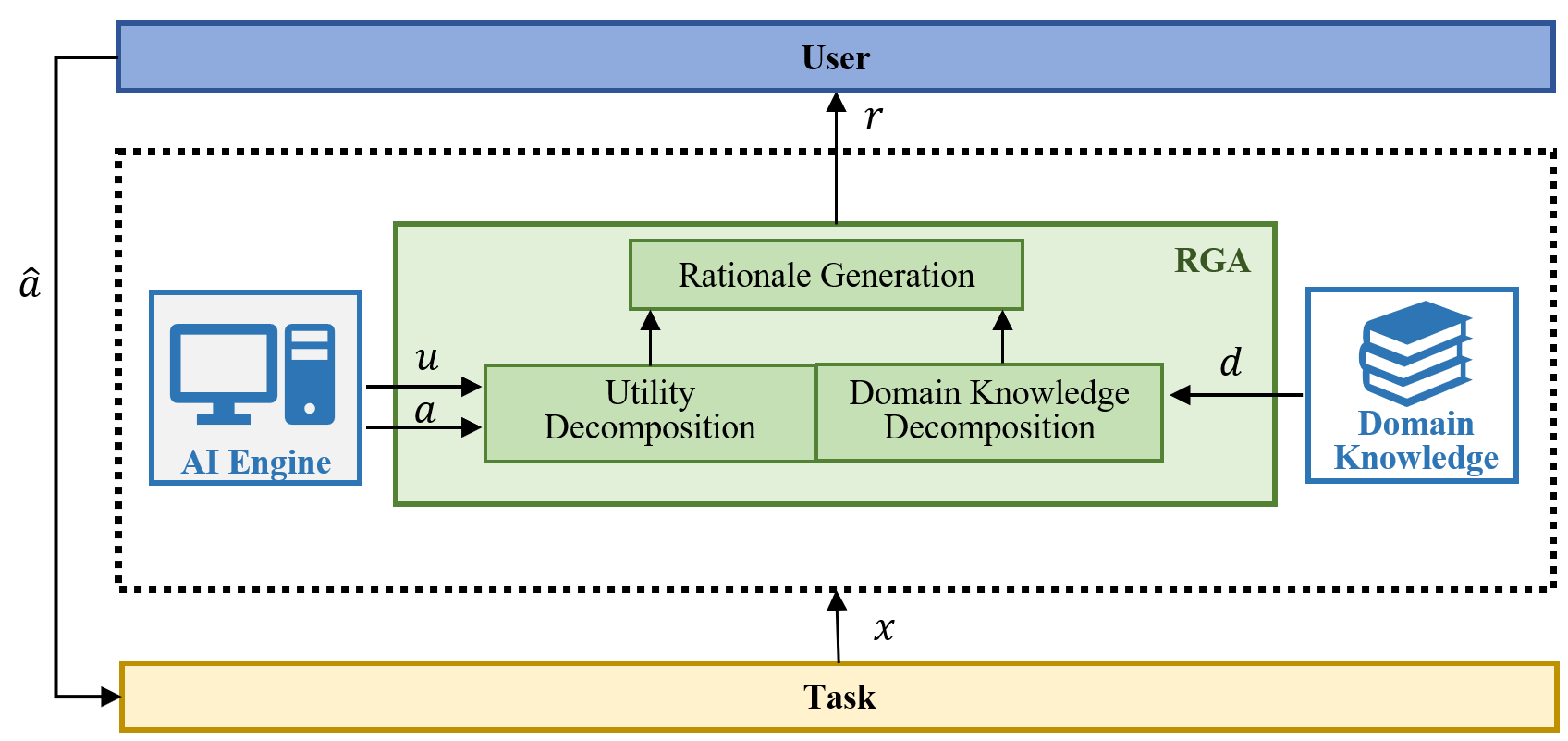}
  \caption{System overview diagram showing the interaction between a user, AI agent and RGA, in which $x$ represents state, $a$ represents action, $u$ represents utility function, $d$ represents domain knowledge and $r$ represents rationale. }~\label{fig:flowchart}
\end{figure}

XAI systems can be used to aid users in a decision-making process, such as medical diagnosis \cite{kermany2018identifying}.  In this work, we introduce the Rationale Generating Algorithm (RGA), which provides the user with rationales designed to aid the user's decision-making process while also increasing their \textit{understanding} of the underlying task domain. Figure \ref{fig:flowchart} presents an overview of the RGA pipeline.  For a given task, we assume an expert AI Engine is available that takes in the current task state $x$, and outputs a recommended action $a$ and its associated utility function $u$\footnote{The presented variant of RGA requires a normalized utility function to generate rationales.  Our objective in this work is to evaluate the effect of rationales on user performance, thus we did not focus on developing a fully model-agnostic rationale generation technique.  In general, approaches such as \cite{ehsan2019automated} can be used to generate rationales for models that do not incorporate a utility function}. In context of this work, a utility function is defined as a weighted sum of all variables used to decide an output. Within RGA, the utility function is decomposed to identify the \textit{most significant factors} contributing to the selection of $a$ over alternate actions.  These factors take the form of variables, typically not very interpretable in their original form (e.g. `PassedPawns $=0.7$').  Additional factors that support decision making may also be obtained from provided expert domain knowledge (see below).  While a utility function may be made up of dozens of variables, RGA selects only the top $k$ factors, which are then used to generate a human-readable rationale. Note that \cite{fox2017explainable} has shown that in the context of an explainable planner, generating justifications to explain both good and bad action choices leads to a more robust explainable system.  Motivated by this, RGA is able to generate rationales for both positively and negatively contributing actions.

\begin{algorithm}[t] 
\caption{generateExp(~$u$,~$a$,~$d$)}  
\textbf{Input} {:~$u$ - utility function, ~$a$ - selected action, ~$d$- domain knowledge} \\ \smallskip
\textbf{Output} {:~$r$ - rationale} \medskip
\begin{algorithmic} [1]
\STATE $P$=[$name_p$:\{\}, $w_p$:\{\}]
\STATE $D$=[$name_d$:\{\}, $w_d$:\{\}] \medskip
\STATE $P$= decomposeUtility($u$)     \medskip
\IF{~$d$ is not \O}
\STATE $D$ = decomposeDomainKnowledge($d$)
\ENDIF \medskip
\STATE $Factors$ = $P \cup D$
\STATE $Factors$.sortByWeight() 
\FOR{$f$ in $Factors$[$i:k$]}
\IF{~isPositive($f$)}
\STATE $r$.append(genPos($f$.name, ~$a$))
\ELSE
\STATE $r$.append(genNeg($f$.name, ~$a$))
\ENDIF
\ENDFOR
\RETURN $r$
\end{algorithmic}
\end{algorithm}

Algorithm 1 further details the manner in which RGA is implemented and a final rationale is chosen.  Variables $P$ and $D$ (lines 1-2) store the name of each decision factor contributing to the selection of a given action, and its associated weight for utility-based ($w_p$) and domain-based ($w_d$) knowledge, respectively. Given the utility function $u$ used to select the given action $a$, RGA first decomposes $u$ to obtain the list of all factors involved in the determination of the action ($name_p$) and their contributing normalized weight ($w_p$).  The resulting list is stored in $P$ (line 3).  

RGA next optionally processes the domain knowledge, if it is available (lines 4-6).  We define domain knowledge as any externally available data, typically encoded a priori by a domain expert \cite{sutcliffe2016domain}. The domain knowledge is encoded as a supplemental utility function $d$ and used to supplement RGA with information that is not encoded in $u$ but might be helpful to include in a rationale to aid human understanding.  For example, within the Chess domain we found that the default $u$ generated by our game engine did not include a variable for 'Checkmate' (an important winning condition in chess), whereas providing information about a checkmate within a rationale would likely be helpful to the user.  As a result, we include the concept of domain knowledge, and in our experimental section compare RGA performance with and without domain knowledge. If domain knowledge is present, RGA decomposes it similarly to the utility function into a list of factor names $name_d$ and their associated weights $w_d$ (line 5).

Once decision factors from the utility function and the domain knowledge are identified, the weighted lists for both sets of factors are merged and sorted by weight in descending order (lines 7-8). The top $k$ factors are then used to generate a rationale, with appropriate wording being dependent on whether the factor positively or negatively contributes to the task objective. For each of the $k$ features that aid in task completion, a positive rationale is scripted (e.g.'A boat can help cross the river because it floats on water'). If the top $k$ factors represent negative contributions to the objective, then a negative rationale is scripted (e.g. 'A car cannot help cross the river because it does not float on water').


Once a rationale is returned by RGA, we display it to the user along with the recommended action $a$ selected by the AI Engine.  The user may then leverage the rationale, and their own knowledge of the domain, to decide whether to perform $a$ or to select a different action. As shown in Figure \ref{fig:flowchart}, the user's final selected action $\hat a$, which may be the same or different from $a$, is then applied back to the task domain.

\section{RGA and Chess}
To measure the effect of human-understandable rationales on task performance, we apply RGA to the game of chess, specifically focusing our attention on endgame configurations, defined by no more than 12 pieces on the board. Chess is a good application for RGA as the utility function for chess is complex in nature, involving many parameters, and the game environment continuously changes over time. We focus specifically on endgame scenarios in which decision making is crucial to an outcome, and there is a relatively small amount of moves left. With endgame scenarios, we are able to analyze the effects of RGA through a measured experimental design. 

For this research, we use the the utility function from the open source Chess AI Engine, Stockfish \cite{Stockfish}, and utilize RGA to generate human-understandable rationale for both an optimal and any non optimal moves taken. This section discusses RGA within the chess domain. 


\begin{figure}
\centering
 \includegraphics[width=0.9\columnwidth]{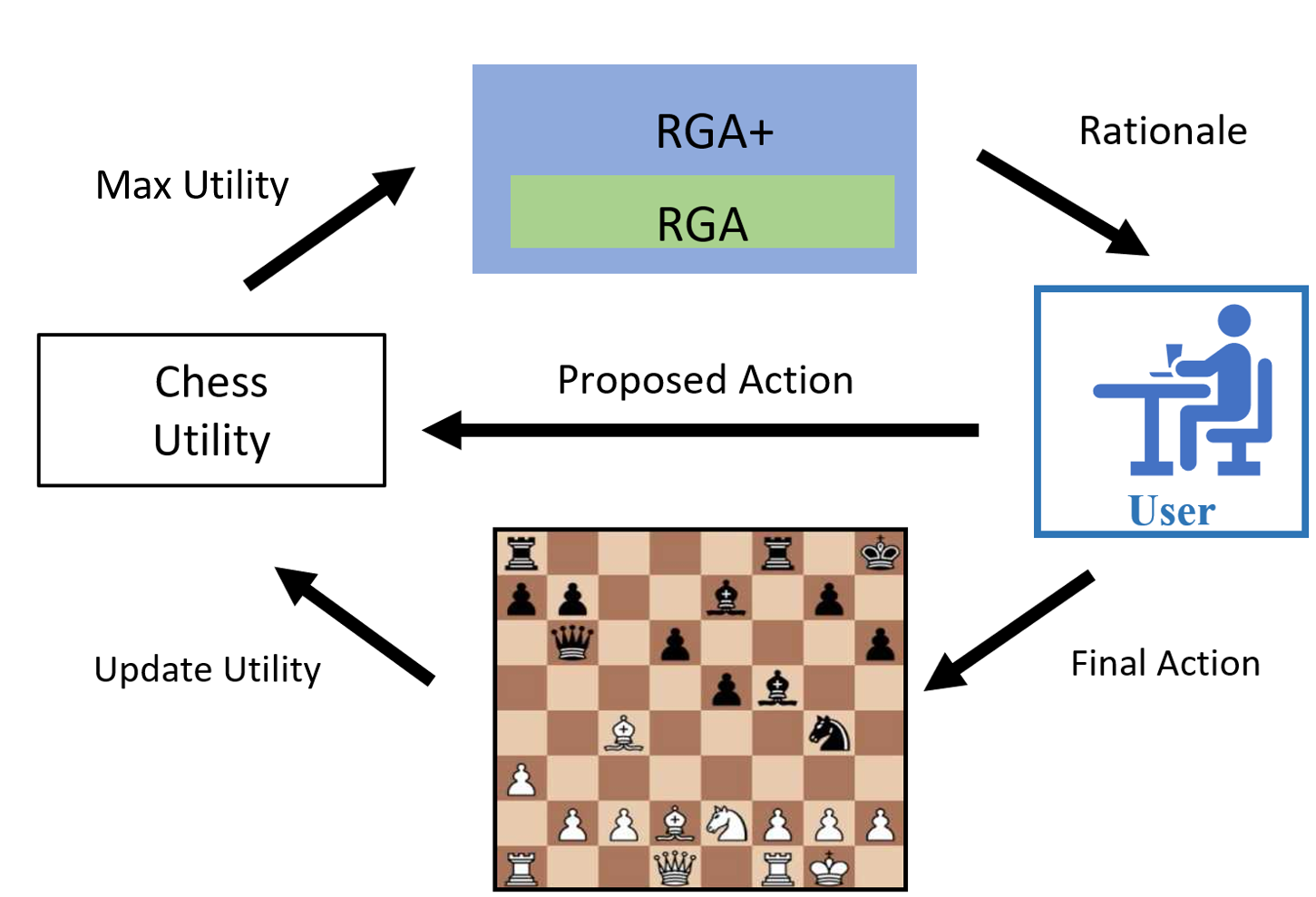}
 \caption{An overview of the application of RGA to the domain of chess, highlighting the two variants of RGA.  }~\label{fig:chess_RGA}
\end{figure}

\begin{figure*}[ht!]
\centering
\begin{subfigure}[b]{0.45\textwidth}
  \centering
  \includegraphics[width=\textwidth]{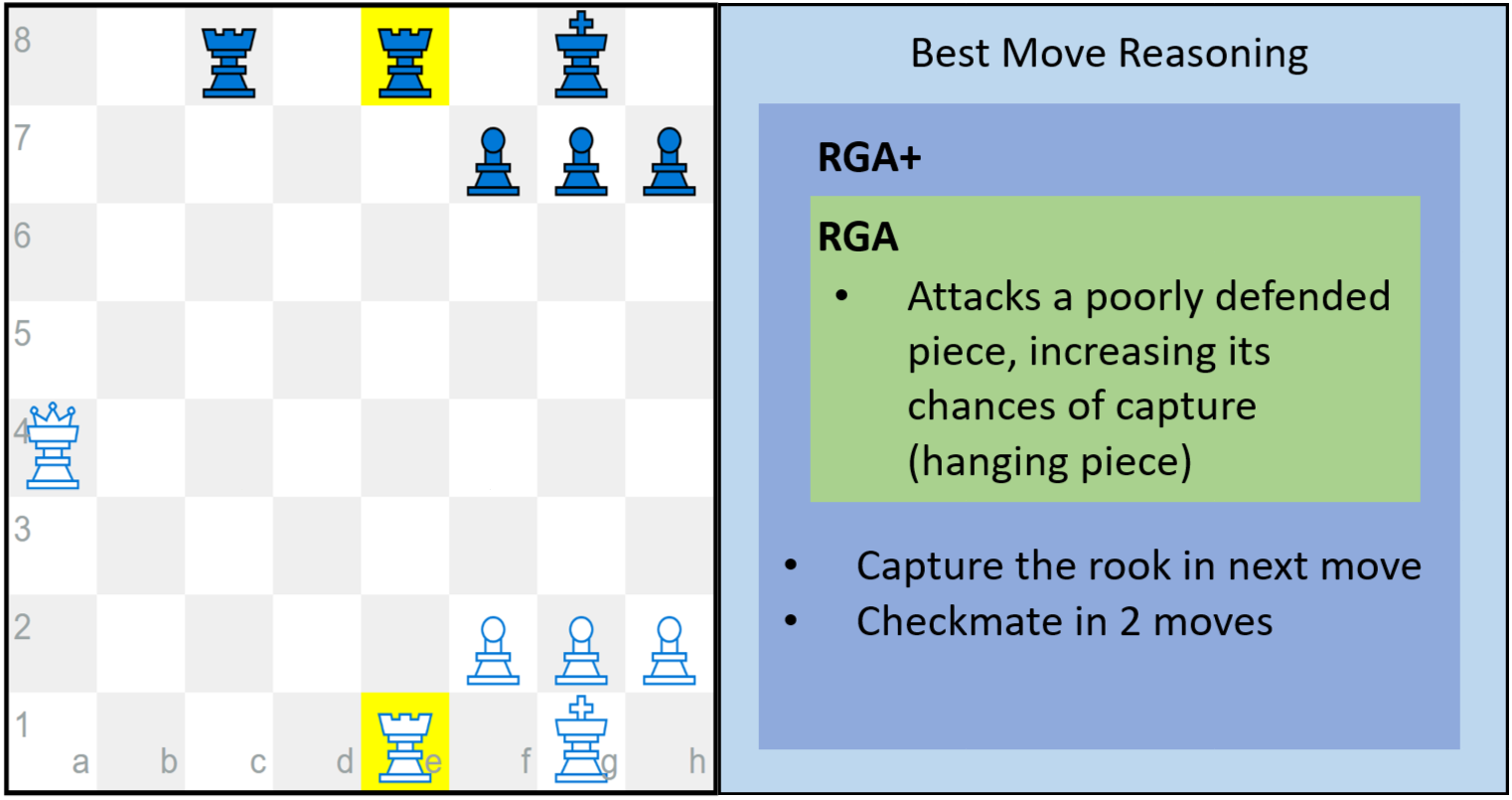}
  \caption[]%
  {{}}
\end{subfigure}\quad
\begin{subfigure}[b]{0.45\textwidth}
  \centering
  \includegraphics[width=\textwidth]{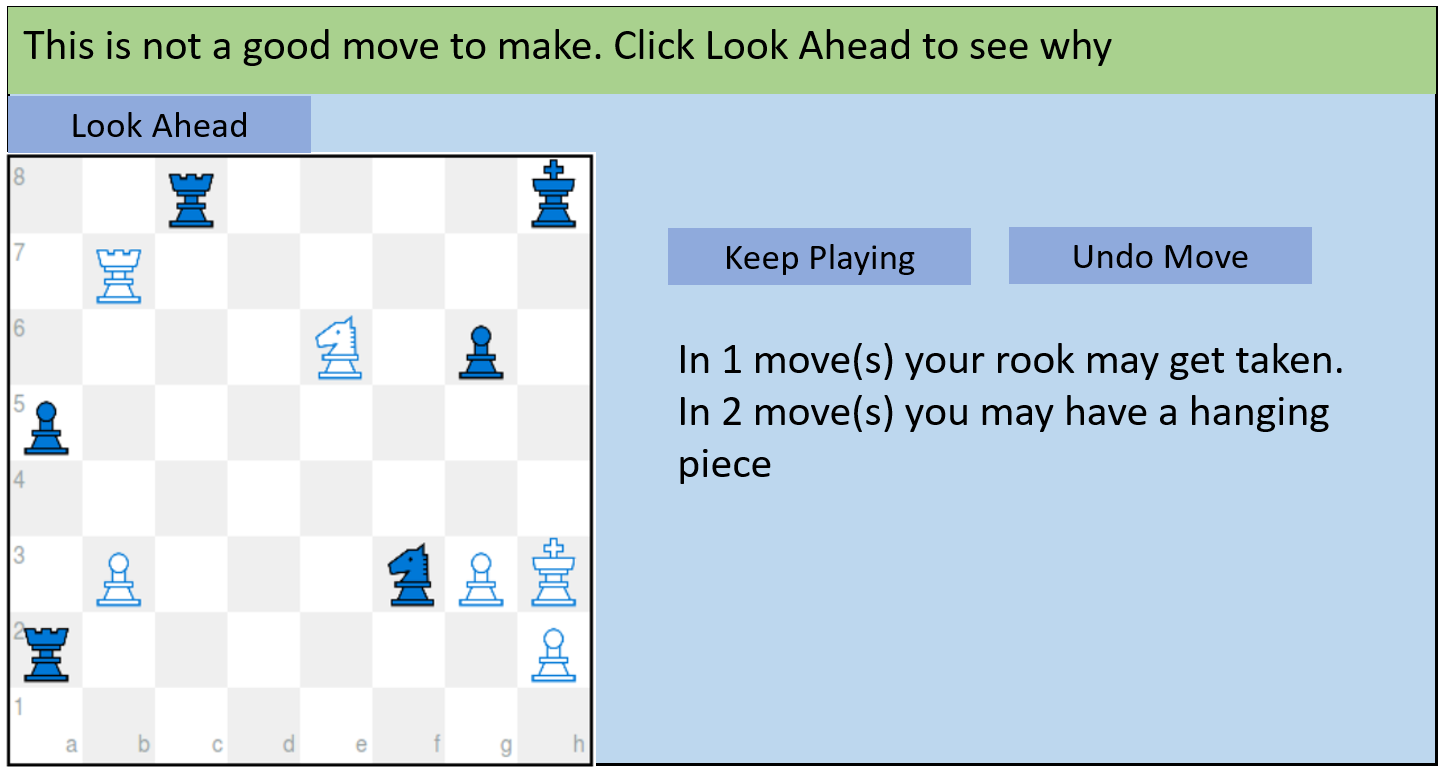}
    \caption[]%
     {{}}
     \end{subfigure}
\caption{Example rationales generated for a particular board configuration where (a) represents a best-move explanation and (b) represents a non-optimal move explanation.}~\label{fig:exampleExps}
\end{figure*}

\subsection{Chess Utility Function}
The utility function from Stockfish includes over 100 utility factors that are combined in a multi-level hierarchical fashion. While all of these factors can be crucial to explaining different moves within a chess board configuration, only a subset of these main factors are useful for explaining moves in context of endgame configurations \cite{bain1994learning}. The relevant endgame utility factors are determined after sorting all factors by maximum weight (line 8 of Algorithm 1). They include but are not limited to: 'Mobility', 'KingDanger', 'King', 'HangingPiece', 'PawnPromotion',  and 'Passed'. 

As a brief summary, 'Mobility' refers to the number of legal moves a player has for a given position--the more choices a player has, the stronger their position on the board. Also for endgames, the king is a powerful piece and is preferred to keep in the center of the board due to its limited range; 'KingDanger' is the highest weighted feature that affects the 'King' score. Additionally, 'Threats' play a huge role in the outcome of an endgame, and are primarily represented by crucially attacking pieces such as rooks and kings, and by 'HangingPiece' which refers to weak enemy pieces not defended by the opponent. 'Passed' refers to the concept of 'Passed Pawns' which checks to see if a pawn can continue to safely advance to the other side of the board. 'Passed', includes the feature 'PawnPromotion' which details when a pawn has reached the opponent's top rank, allowing the player to switch its pawn out for a queen, rook, bishop or knight. These described utility factors represent the utility factors that we deemed appropriate from the Stockfish utility function, based on their weights, to use in justifying a non-optimal or optimal move choice. 

\subsection{Data Pre-Processing}
To meet RGA's requirement of having a normalized utility function input, we standardize the Stockfish utility function using Z-score scaling. For Z-score scaling, we perform a heuristic ad-hoc analysis by collecting average standard deviations and means for each relevant factor over several game configurations. Specifically, we collect 120 game configurations using a random FEN generator. To ensure completeness, we represent an equal distribution of game configurations within the range of 2-32 pieces. We started with 30 random game configurations, and doubled them twice, until the change in both average standard deviation and mean for each factor was negligible. 

 
\subsection{RGA In Chess} 
In the context of chess, RGA generates human understandable rationales with the objective of checkmating the opponent. As portrayed in Figure \ref{fig:flowchart}, the Stockfish AI engine generates a recommended action for the current board state~$x$, and its associated utility $u$. In specific relation to chess, Figure \ref{fig:chess_RGA} depicts the overall interaction between the Chess Engine, RGA, and the user. For each given board configuration, all relevant factors, actions and utilities are updated before generating a rationale. These utilities are sorted to find the highest utility $u_{best}$ and corresponding optimal move $a_{best}$, which RGA uses to generate a rationale to the user to justify the optimal move he/she should take. Additionally, we detect non-optimal moves by comparing the user's proposed action to the set of all possible actions $A$ for the given board configuration; if the proposed action falls in the bottom 1/3 of $A$, then we use the factors in $Factors$ (line 7 of Algorithm 1) to generate a cautionary rationale justifying why a user should not make the proposed action. The user can ultimately decide upon the final action, considering or disregarding the rationales produced by RGA. Figure \ref{fig:exampleExps} provides an example of a best-move and a non-optimal move explanation using both a possible Stockfish and domain-knowledge factor.


\subsection{Domain Knowledge Factors}
In the application of chess, we use~$domainKnowledge()$ to adds three important criteria that the Stockfish utility function does not explicitly represent: (1) explicit piece capture on next move, (2) check on next move, (3) checkmate on next or subsequent move. Considering the objective of chess, we weigh these additional domain-knowledge factors higher than those from the utility function. Shown in Figure~\ref{fig:chess_RGA}, we distinguish RGA+ to be a superset of RGA. The overarching RGA+ denotes rationales that are reasoned from both domain knowledge and the chess utility function, whereas the RGA requires a utility function for rationale generation, but leaves domain knowledge inputs as optional to domain experts. 


\section{Experimental Design}
To evaluate the effectiveness of RGA in improving task performance, we conducted a four-way between-subjects user study in which participants took part in chess gaming sessions over three consecutive days.  We selected a multi-day study design to ensure observation of longer-term learning effects.  Given the complexity of chess, which requires an estimated 10 years \cite{simon1988skill} or 5000 hours \cite{charness2005role} to master, we conducted our study using only simplified game scenarios consisting of \textit{end games}.  

\begin{figure*}[t]
\centering
\begin{subfigure}[b]{0.23\textwidth}
  \centering
  \includegraphics[width=\textwidth]{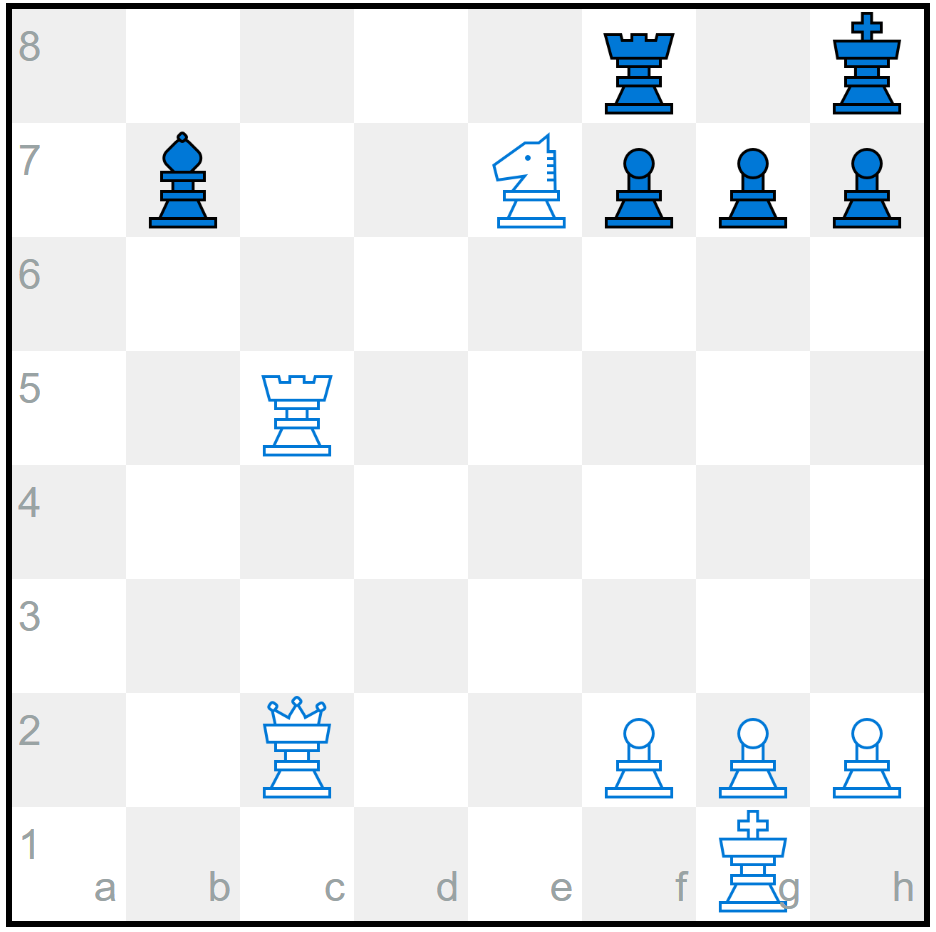}
  \caption[]{None}
\end{subfigure}\quad
\begin{subfigure}[b]{0.23\textwidth}
  \centering
  \includegraphics[width=\textwidth]{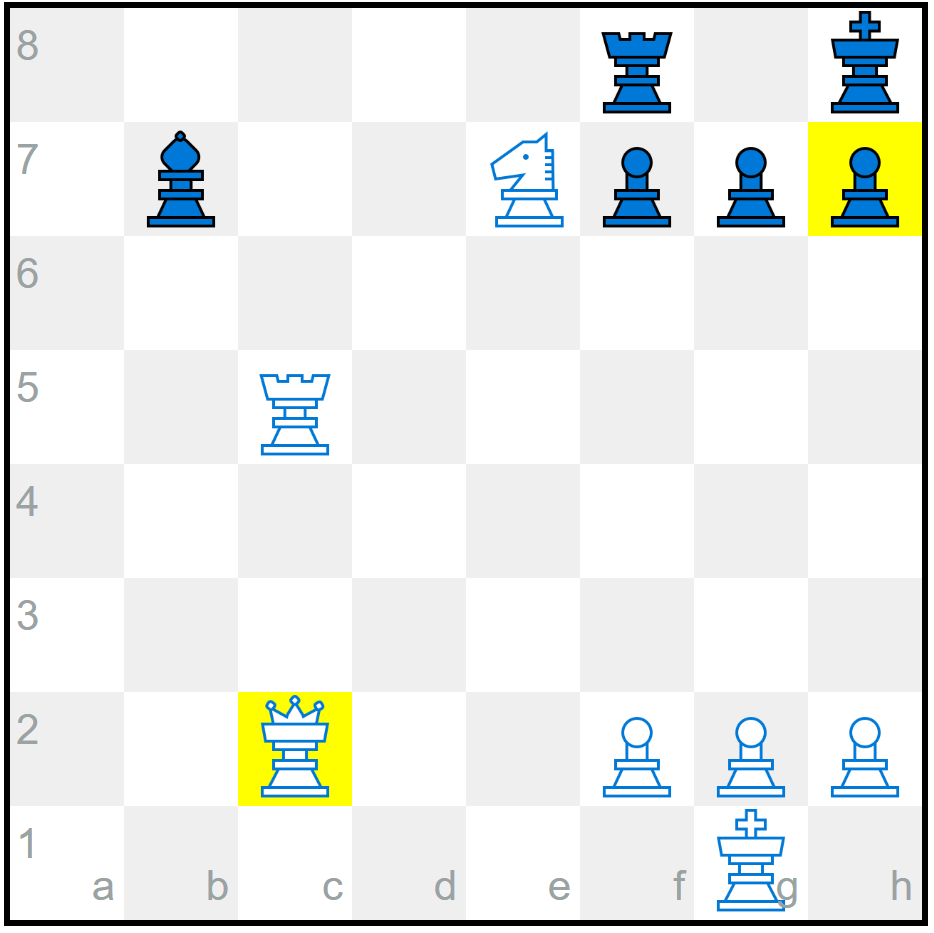}
    \caption[]{Hints}
     \end{subfigure}\quad
\begin{subfigure}[b]{0.43\textwidth}
  \centering
  \includegraphics[width=\textwidth]{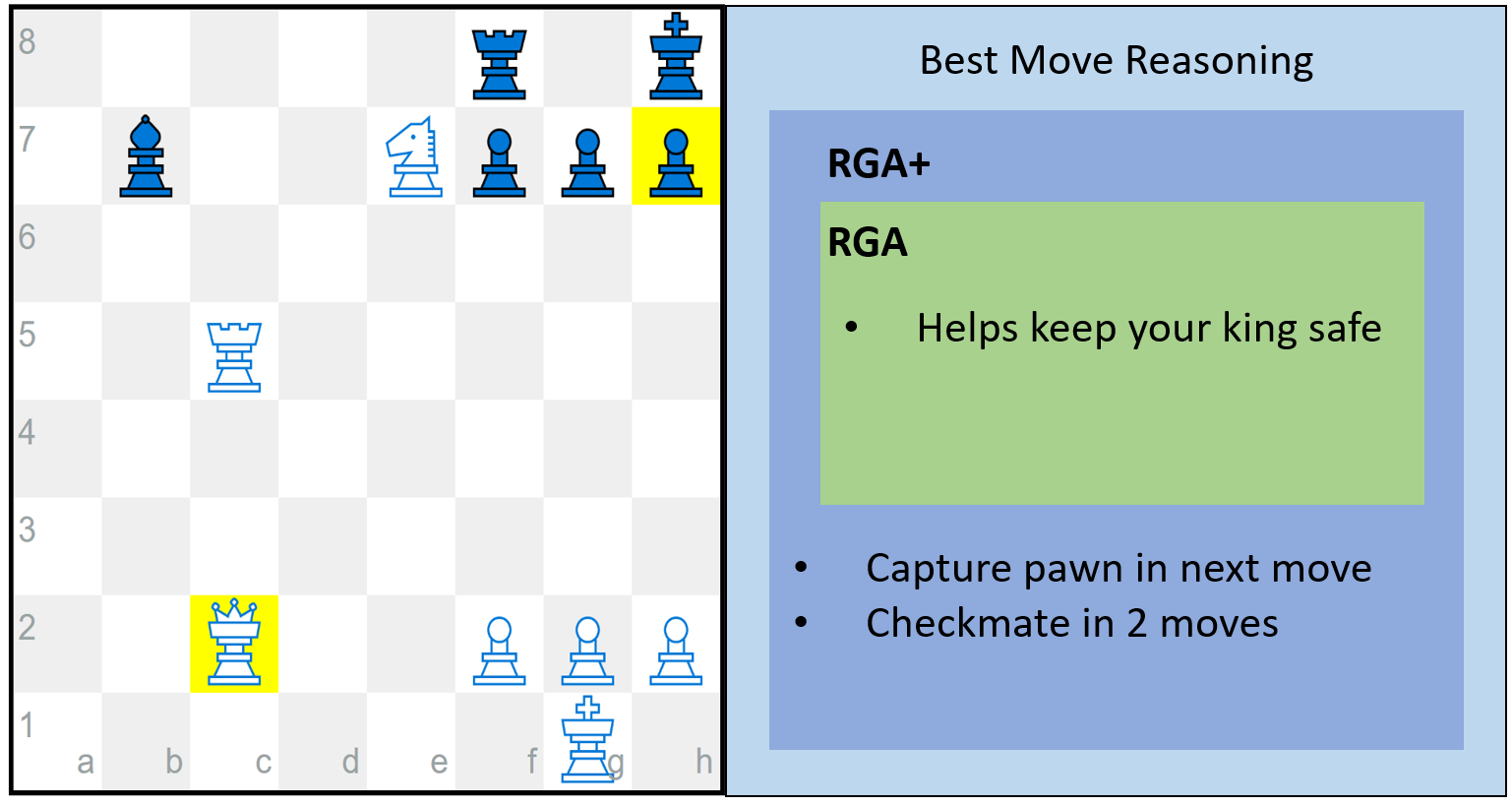}
    \caption[]{RGA (green) and RGA+ (blue)}
     \end{subfigure}
\caption{Given a set game configuration, the three chessboards show the different user interfaces for the four experimental conditions where (a) represents the 'None' cohort, (b) represents 'Hints' and (c) represents the 'RGA+ and RGA' cohort.}
\label{fig:gameConfigurations}
\end{figure*}

The study design consisted of the following four study conditions, which determine what guidance was provided to the participant: 

\begin{flushitemize}
\item \textbf{None (baseline)}: The player receives no hints or rationales.  This condition is equivalent to practicing chess independently with no guidance. (Figure \ref{fig:gameConfigurations}(a))

\item \textbf{Hints (baseline)}: The player receives a visual hint highlighting in color the best currently available move, as determined by the game engine utility function.  No textual rationales are provided beyond the highlighting.  This condition is equivalent to the hints system available in the vast majority of online and app-based chess programs.  (Figure \ref{fig:gameConfigurations}(b))

\item \textbf{RGA}: The player receives a visual hint highlighting in color the best currently available move (as in Hints). Additionally, the system displays a textual rationale based the Stockfish utility function only.  (Figure \ref{fig:gameConfigurations}(c))

\item \textbf{RGA+}: The player receives a visual hint highlighting in color the best currently available move (as in Hints).  Additionally, the system displays a textual rationale based on both the Stockfish utility function and domain knowledge.  (Figure \ref{fig:gameConfigurations}(c))
\end{flushitemize}

The sections below further detail our evaluation metrics, study hypotheses, participant recruitment method, and study design.

\subsection{Metrics}
Each chess session consisted of \textit{diagnostic games}, during which participants were evaluated on their performance and received no suggestions or hints, and \textit{instructional games}, during which participants received guidance according to their assigned study condition.  During diagnostic games, the following metrics were used to evaluate performance:
\begin{flushitemize}
\item \textbf{Win Percentage (\textit{Win\%}):} a metric commonly used in sports that takes into account the number of \textit{wins}, \textit{losses} and \textit{ties}.  In our domain, we additionally account for cases in which the maximum number of allowed moves has been reached\footnote{We limit the total number of moves to prevent novice players from moving around the board indefinitely.}, \textit{maxmoves}, which are weighted the same as ties.  The final win percentage is calculated as:  
\begin{equation}
   {Win\%}= \frac{wins + 0.5 \cdot ties + 0.5 \cdot maxmoves}{wins + ties + losses + maxmoves}
   \end{equation}
\item \textbf{Percentile Rank (\textit{Percentile})}: a metric used to measure the distribution of scores below a certain percentage. In our domain, the distribution of scores is the move rating of all possible moves for a given board configuration. The move ratings are calculated by Stockfish. For each board configuration, we use the following formula to calculate the percentile rank of a chosen move ~$k$ with ~$N$ corresponding to all possible moves:
\begin{equation}
   {Percentile Rank}= \frac{100\cdot(\left(k-1\right))}{\left(N-1\right)}
   \end{equation}
\end{flushitemize}

Additionally, at the end of each study day, participants were given a short post-session questionnaire from which we obtain the following metric:

\begin{flushitemize}
\item \textbf{Perceived Performance (\textit{SelfEval})}: a metric that seeks to capture the participants' self-reported perceived progress toward learning chess. Perceived performance is measured using a 5-Point Likert Scale rating based on the question `Do you believe your performance improved this session?' (1 = Strongly disagree, 5 = Strongly agree')
\end{flushitemize}

\subsection{Hypotheses}
We formulate the following hypotheses on the ability of interpretable rationales to improve participants' chess performance defined by the \textit{Win\%} and \textit{Percentile Rank} metrics:
\begin{flushitemize}
\item \textbf{H1a}: Participants who received only utility-based rationales (RGA) will perform better than those who received no guidance (None).
\item \textbf{H1b}: Participants who received only utility-based rationales (RGA) will perform better than participants who received only suggestions (Hints). 
\item \textbf{H1c}: Participants who received rationales that incorporate domain knowledge (RGA+) will perform better than participants who received no guidance (None).
\item \textbf{H1d}: Participants who received rationales that incorporate domain knowledge (RGA+) will perform better participants who received only suggestions (Hints).
\item \textbf{H1e}: Participants who received rationales that incorporate domain knowledge (RGA+) will perform better than participants who received only utility-based rationales (RGA).
\end{flushitemize}

Additionally, we hypothesize that participants' measure of their own perceived performance, evaluated by the \textit{SelfEval} metric, will follow similar trends as above.  Specifically: 
\begin{flushitemize}
\item \textbf{H2a}: Participants who received only utility-based rationales (RGA) will have higher perceived performance ratings than those who received no guidance (None).
\item \textbf{H2b}: Participants who received only utility-based rationales (RGA) will have higher perceived performance ratings than those who received only suggestions (Hints).
\item \textbf{H2c}: Participants who received rationales that incorporate domain knowledge (RGA+) will have higher perceived performance than participants who received no guidance (None)
\item \textbf{H1d}: Participants who received rationales that incorporate domain knowledge (RGA+) will have higher perceived performance than who received only suggestions (Hints).
\item \textbf{H2e}: Participants who received rationales that incorporate domain knowledge (RGA+) will have higher perceived performance ratings than those who received only utility-based rationales (RGA).
\end{flushitemize}

\subsection{Participants}
We recruited 68 participants from Amazon's Mechanical Turk.  Participants were required to demonstrate basic knowledge of chess by passing a short test verifying the rules of the game (e.g.'Which piece can move straight forward, but captures diagonally?'). Participants were also required to participate three days in a row, and to not already be expert players. Eight players were removed from the study for either not participating for three days or for winning every game (suggesting that they were experts to begin with). The final set of participants included 60 individuals (44 males and 16 females), who ranged in age from 18 to 54 (6 between 18-24 years, 31 between 25-34 years, 18 between 35-44, and 3 between 45-54 years). Participants were randomly assigned to one of the four study conditions. Each daily session took approximately 15-20 minutes, and participants were compensated \$2.00, \$4.00, and \$6.00 on days 1, 2 and 3, respectively.

\subsection{Study Design}
The study consisted of three sessions performed on three consecutive days. In each session, participants played 9 games of chess: three diagnostic games, followed by three instructional games, followed by three more diagnostic games. The use of diagnostic games at the beginning and end of each session enabled us to study participant performance both across and within sessions\footnote{Our analysis showed no significant trends for within-session performance differences, likely due to the short duration of the sessions. However, we do observe significant learning effects across sessions, as discussed in the Section \ref{sec:results}.}.

The participant always played white, and the opponent black pieces were controlled by the Stockfish Engine AI, which always played optimally. For each board, the optimal number of player moves needed to win,~$O$, was determined by the Stockfish Engine, and participants were limited to 10 moves during the game. Starting board configurations were obtained from a popular online learning website (\url{https://lichess.org/}), selected such that~$O<5$. All participants received the same boards in the same order to ensure uniformity; each board was unique and did not repeat. Players were allowed to make any legal chess move, and each game consisted of an average of 6 moves (SD=2.62). As a result, participants in the Hints and RGA conditions received approximately 18 move suggestions per day on average. Furthermore, participants in the RGA condition received one rationale per move suggestion, receiving 18 rationales per day. Participants in the RGA+ conditions occasionally received an additional rationale per move suggestion to denote a possible checkmate in less than three moves.

\begin{figure}
\centering
  \includegraphics[width=0.9\columnwidth]{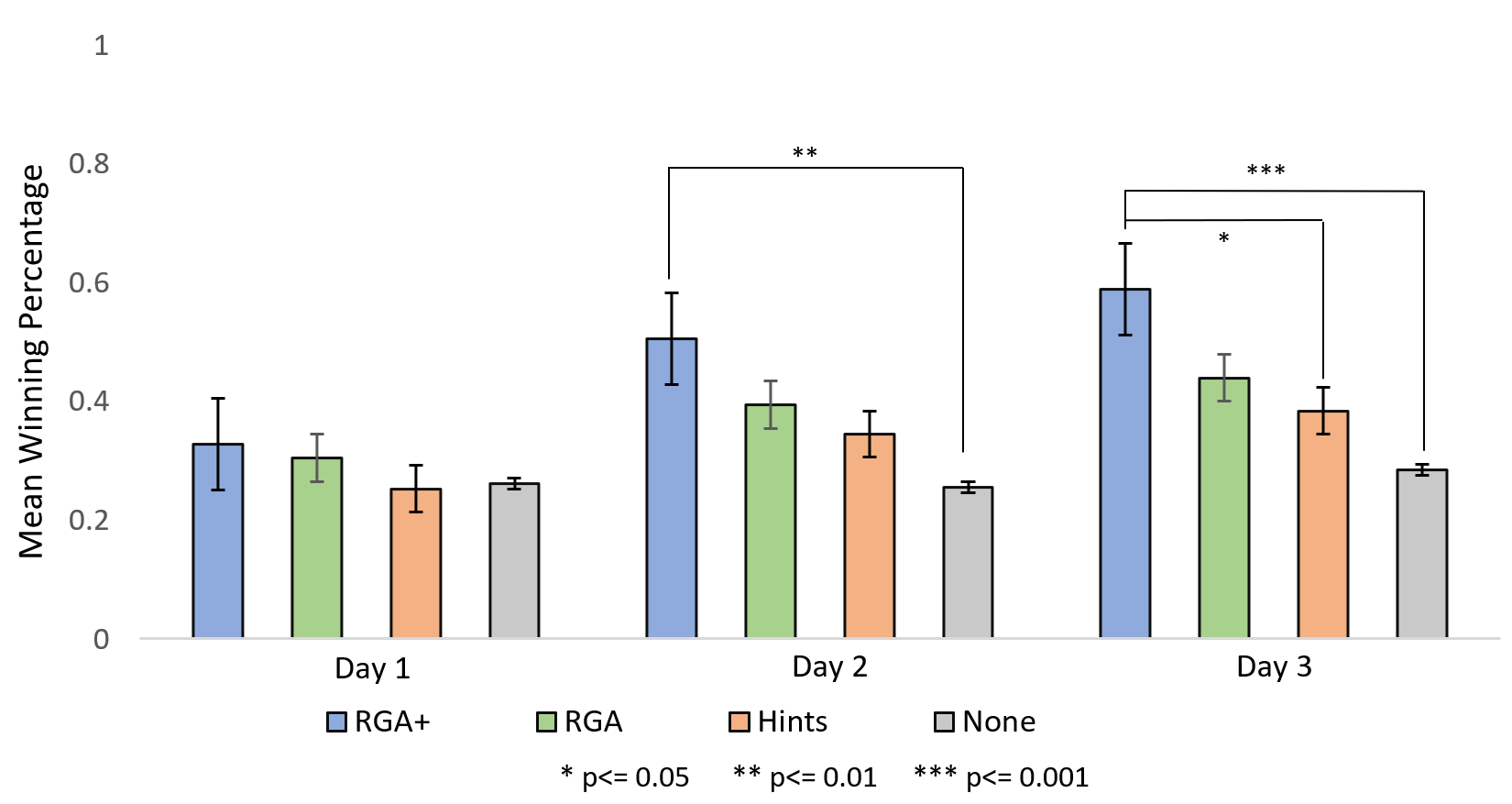}
  \caption{ Average \textit{Win\%} of the experimental conditions.}~\label{fig:win_data}
\end{figure}
\begin{figure}
\centering
  \includegraphics[width=0.9\columnwidth]{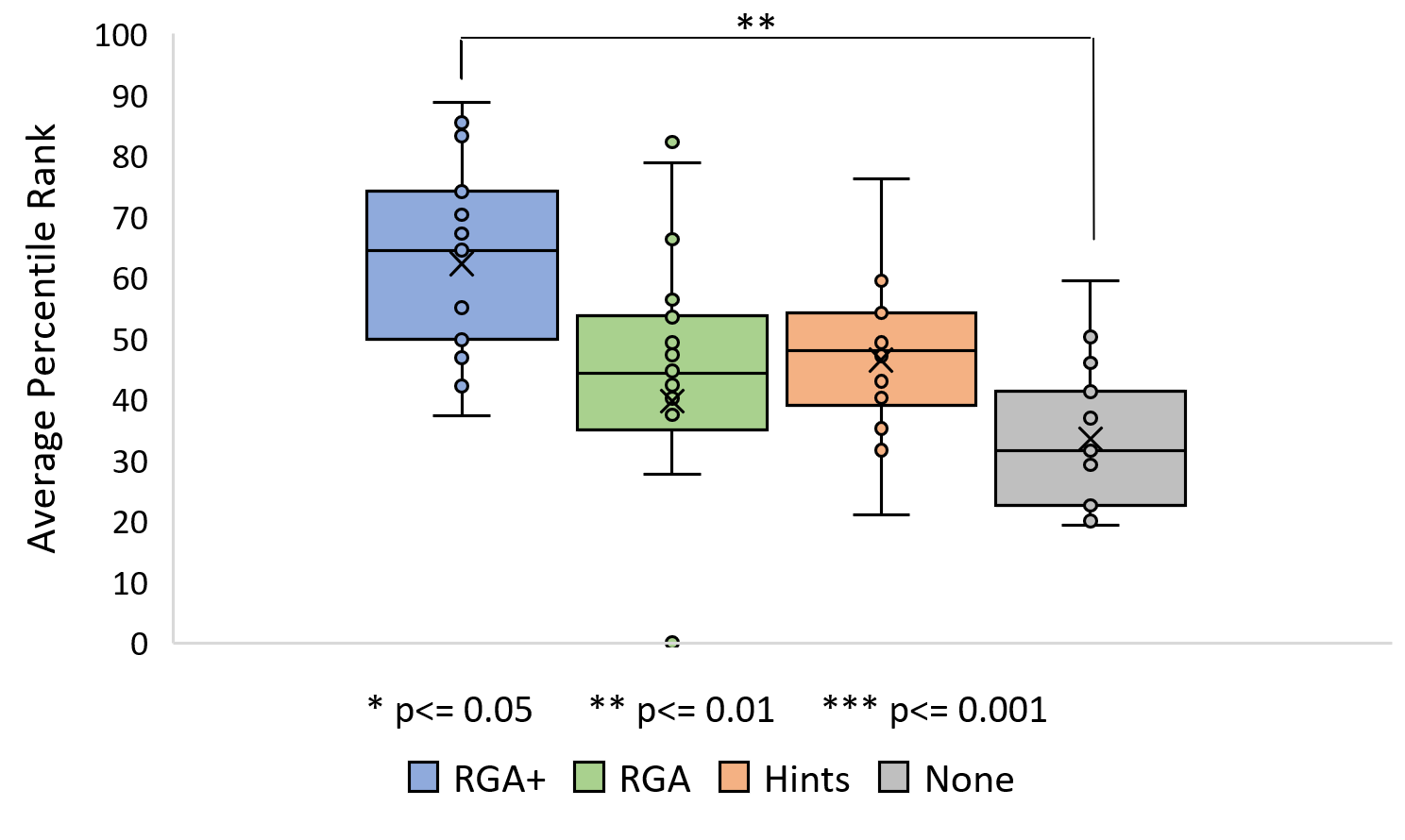}
  \caption{Average \textit{PercentileRank} ratings of moves made by participants across all days.}~\label{fig:rank_data}
\end{figure}

\section{Results}
\label{sec:results}
The participant performance data followed a normal distribution; as a result, we used ANOVA with a Tukey HSD post-hoc test to evaluate statistical significance across the experimental conditions with respect to the \textit{Win\%} and \textit{PercentileRank} metrics. Additionally, we conducted a Mann-Whitney U test to analyze the Likert scale data for the \textit{SelfEval} metric.

\subsection{Participant Performance}
Figure \ref{fig:win_data} presents the average participant win percentage (\textit{Win\%}) for each study condition over the three days. We observe that while no statistical differences are observed between conditions on the first day, the differences in performance grow on subsequent days. In particular we see the greatest rate of overall task improvement from Day 1 to Day 3 for the RGA+ condition.
Our results indicate a correlation between the amount of explanation participants are given and their \textit{Win\%}, with more justifications leading to more wins. 
This is further supported by the results in Figure \ref{fig:rank_data}, which present the percentile rank (\textit{PercentileRank}) of each participant's average move--with a percentile rank of 100 denoting the most optimal move and percentile rank of 0 denoting the least optimal move. Similar to Figure \ref{fig:win_data}, we observe a correlation between the \textit{PercentileRank} of move ratings and the amount of justifications provided, with the upper quartile for \textit{PercentileRank} being higher with more explanations. 

Our statistical analysis shows that for both \textit{Win\%} and for \textit{PercentileRank}, participants in 'RGA' did not perform statistically better than those in 'None', \textbf{not validating H1a}, nor statistically better those in 'Hints', \textbf{not validating H1b}. We do observe that 'RGA+' participants performed significantly better than 'None' participants \textbf{validating H1c}.  However, with respect to the 'Hints' condition, 'RGA+' had statistically better \textit{Win\%} performance but not \textit{PercentileRank}, therefore only \textbf{partially validating H1d}. Finally, we see no significant difference between 'RGA+' and 'RGA' conditions in this study, \textbf{not validating H1e}.

In summary, these results indicate that humanly interpretable rationales \textit{can} improve task performance, as long as the rationales fit a complete representation of the domain. Our results show that gathering both utility-based features and additional domain knowledge features (not represented by the utility) can accomplish completeness. It is also important to note that for task improvement, a decision not only needs to be interpretable, but more importantly \textit{humanly} interpretable by accompanying rationales.
The 'Hints' conditions also provided interpretability by highlighting the best move, but the lack of accompanying rationales may be the cause of why no significance was seen between 'Hints' and 'None'. Furthermore the inclusion of domain knowledge in RGA+ significantly improved participant \textit{Win\%} over the baseline conditions, compared to RGA, validating the need for a more complete domain representation. However, for this reason, it is interesting that no \textit{Win\%} significance was seen between 'RGA+' and 'RGA', but in seeing a trend of increasing difference between 'RGA+' and 'RGA', we suspect that over a longer period of time, a visible significance would be seen. 

\subsection{Participant Perceived Performance}
\begin{figure}
\centering
  \includegraphics[width=0.9\columnwidth]{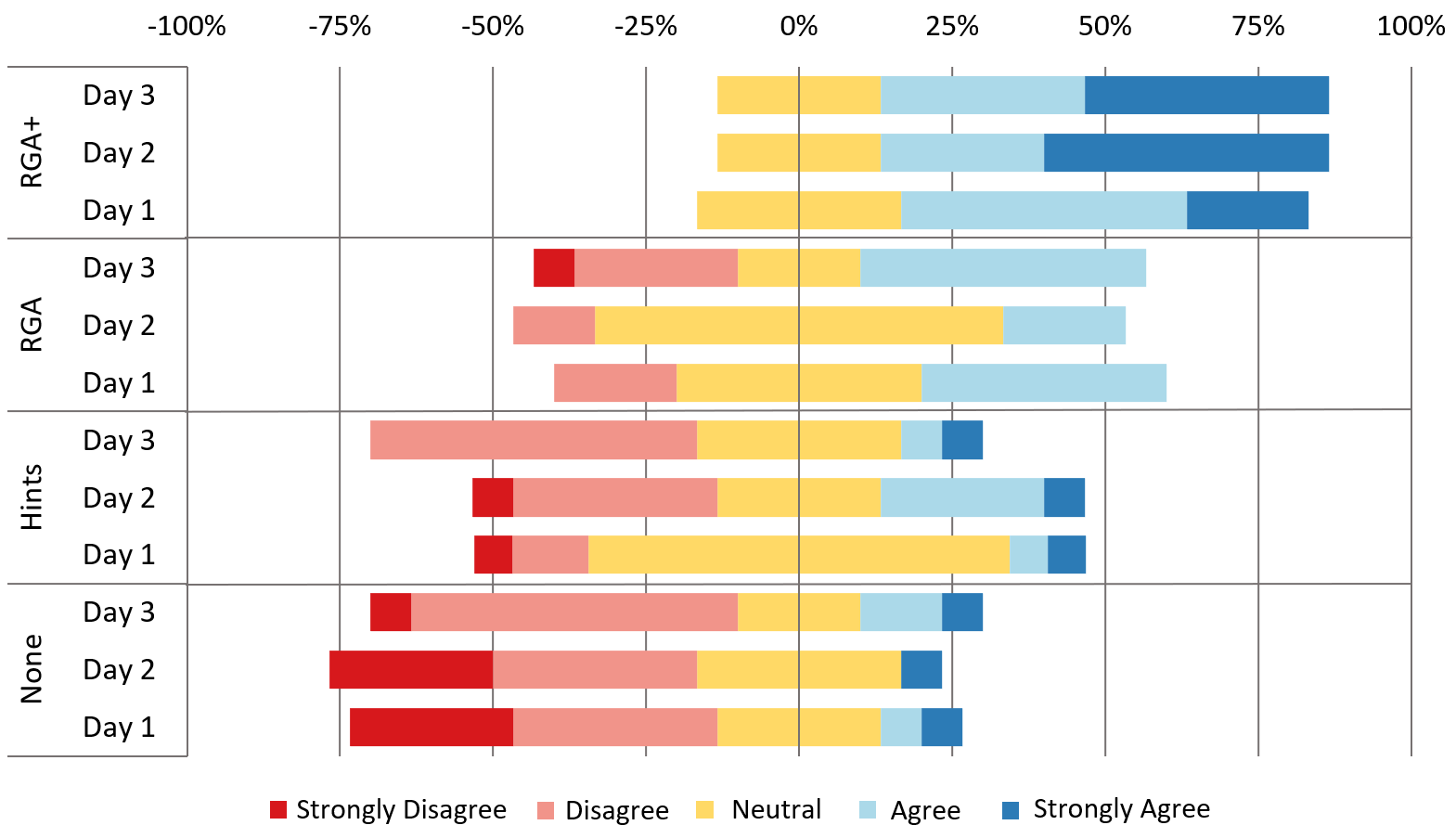}
  \caption{Survey data of participants' \textit{SelfEval} across the experimental conditions.}~\label{fig:likert}
\end{figure}

Figure \ref{fig:likert} presents the perceived performance rating (\textit{SelfEval}) of participants in each experimental condition. The Likert scale data shows that participant groups that received more justifications ('RGA+', 'RGA') had higher ratings of 'Agree' and 'Strongly Agree' than participant groups that received little to no justifications ('Hints', 'None'), showing the value justifications had on \textit{SelfEval}. Furthermore, participant groups that received some justifications ('RGA', 'Hints') had more 'Neutral' ratings than participant groups that were on the extreme ends of the justification spectrum ('RGA+', 'None'), showing higher levels of uncertainty in their \textit{SelfEval}. 

The Mann-Whitney U tests in Table~\ref{tab:Signifs} further detail the specific significance between each experimental condition.  As seen in Table~\ref{tab:Signifs}, the \textit{SelfEval} of 'RGA' participants were not significantly stronger than 'None' participants' ratings, \textbf{not validating H2a}, nor significantly stronger than 'Hint' participant ratings, \textbf{not validating H2b}. 
However, the 'RGA+' participants did have a statistically higher \textit{SelfEval} than 'None' participants and 'Hints' participants on all three days, \textbf{validating H2c} and \textbf{validating H2d} respectively. Additionally, from day two onward, 'RGA+' participants rated their perceived performance higher than those from 'RGA', \textbf{validating H2e}. 

Overall, the \textit{SelfEval} metric data aligns with the performance analysis from Section 6.1, showing that perceived performance ratings are significantly stronger with the presence of humanly interpretable rationales that are domain representative. The results above also portray additional significance not seen with the \textit{Win\%} and \textit{PercentileRank} metrics. Unlike the trend in Section 6.1, \textit{SelfEval} does show a significant difference in performance rating between 'RGA+' and 'RGA', reiterating the importance of holistic domain representation. Interestingly, similar to the analysis from \textit{Win\%} and \textit{PercentileRank}, \textit{SelfEval} also does not portray a significant difference between 'RGA' and 'Hints', implying that rationales from the utility function alone were not different enough from 'Hints'. In Figure \ref{fig:likert}, we see an increasing difference in 'Disagree' ratings and 'Agree' ratings over Day 1 and Day 3 between 'RGA' and 'Hints', implying that over a longer period of observation, a potential significance could be seen. 

\section{Discussion and Conclusions}
In this work, we are the first to explore whether human-interpretable rationales automatically generated based on an AI's internal task representation can be used to not just explain the AI's reasoning, but also enable end users to better understand the task itself, thereby leading to improved user task performance.  Our work introduces the \textit{Rationale-Generating Algorithm} that utilizes utility based computational methods to produce rationales understandable beyond the scope of domain-expert users. To validate RGA, we applied it to the domain of chess and measured human task performance using both qualitative user self-reported data (self-perceived performance ratings) and quantitative performance measures (winning percentages and rank percentiles of the strength of moves played by each participant).  

\begin{center}
\begin{table}
\begin{tabular}{ |p{2cm}||p{1.3cm}|p{1.3cm}|p{1.3cm}|}
\hline
 Conditions& Day 1&Day2&Day3\\
 \hline
 RGA+ vs. RGA   & NS    & $p \le 0.01$ & $p \le 0.05$ \\
 RGA+ vs. Hints& $p \le 0.01$  & $p \le 0.01$   & $p \le 0.001$ \\
 RGA+ vs. None & $p \le 0.001$ & $p \le 0.001$ &  $p \le 0.001$ \\
 RGA vs. Hints    &NS & NS&  NS\\
 RGA vs None& $p \le 0.05$  & $p \le 0.01$ &NS\\
 Hints vs None& NS  & NS   &NS\\
\hline
\end{tabular}
\caption{ Mann-Whitney U test significance values for the \textit{SelfEval} metric }
\label{tab:Signifs}
\end{table}
\end{center}

Our results show that rationales from RGA are effective in improving performance when information from the AI's utility function is combined with additional domain knowledge from an expert. The resulting system was able to statistically significantly improve user performance in chess compared to study participants who practiced the same number of games but did not receive rationales.  Simply showing participants the optimal action without an accompanying rationale did not produce the same results, indicating the importance of interpretable rationales in elucidating the task.  

The presented approach is the first study of how rationales can affect learning.  While it contributes a number of important insights, it is also limited in several ways.  First, RGA is limited to utility-based methods and can not be applied to arbitrary machine learning methods.  Future work in generating rationales for alternate ML representations, such as reinforcement learning discussed in \cite{ehsan2019automated}, should be explored. Exploration of such methodologies can lead to developing a rationale generating system that is model agnostic.  Second, our work does not compare among the many different ways in which rationales can be phrased or structured. It is beneficial to investigate if and how non-verbal explanations and dynamic forms of explanations can provide better humanly understandable explanations. For example, it would be interesting to see whether visual animations of future chess moves help RGA rationales be as effective as RGA+ rationales as well as the effect visual animations have (applied to the Hints and None category) without RGA. Also, additional research is needed to evaluate how to present rationales in the most accessible and interpretable manner based on individual needs. Currently, RGA is developed to help beginner chess players improve their task performance, but it is worth exploring how to build a \textit{learned} RGA that can tailor its explanations to varying levels of expertise. Another important area to investigate is the long term effects of RGA. While RGA has been shown as valuable in improving human task performance in a short period of time, it would be beneficial to see whether these trends are upheld over a longer time frame. Learning when the effects of RGA are minimal and when they are maximum can help establish its best time of usage, as well as measure its broader impact in improving human learning and human task performance.


\begin{acks}
This material is based upon work supported by the NSF Graduate Research Fellowship under Grant No. DGE-1650044, and also in part by NSF IIS 1564080 and ONR N000141612835.
\end{acks}

\bibliographystyle{ACM-Reference-Format}
\bibliography{acmart}


\begin{thebibliography}{34}


\ifx \showCODEN    \undefined \def \showCODEN     #1{\unskip}     \fi
\ifx \showDOI      \undefined \def \showDOI       #1{#1}\fi
\ifx \showISBNx    \undefined \def \showISBNx     #1{\unskip}     \fi
\ifx \showISBNxiii \undefined \def \showISBNxiii  #1{\unskip}     \fi
\ifx \showISSN     \undefined \def \showISSN      #1{\unskip}     \fi
\ifx \showLCCN     \undefined \def \showLCCN      #1{\unskip}     \fi
\ifx \shownote     \undefined \def \shownote      #1{#1}          \fi
\ifx \showarticletitle \undefined \def \showarticletitle #1{#1}   \fi
\ifx \showURL      \undefined \def \showURL       {\relax}        \fi
\providecommand\bibfield[2]{#2}
\providecommand\bibinfo[2]{#2}
\providecommand\natexlab[1]{#1}
\providecommand\showeprint[2][]{arXiv:#2}

\bibitem[\protect\citeauthoryear{Abdul, Vermeulen, Wang, Lim, and
  Kankanhalli}{Abdul et~al\mbox{.}}{2018}]%
        {abdul2018trends}
\bibfield{author}{\bibinfo{person}{Ashraf Abdul}, \bibinfo{person}{Jo
  Vermeulen}, \bibinfo{person}{Danding Wang}, \bibinfo{person}{Brian~Y Lim},
  {and} \bibinfo{person}{Mohan Kankanhalli}.} \bibinfo{year}{2018}\natexlab{}.
\newblock \showarticletitle{Trends and trajectories for explainable,
  accountable and intelligible systems: An hci research agenda}. In
  \bibinfo{booktitle}{\emph{Proceedings of the 2018 CHI conference on human
  factors in computing systems}}. ACM, \bibinfo{pages}{582}.
\newblock


\bibitem[\protect\citeauthoryear{Adadi and Berrada}{Adadi and Berrada}{2018}]%
        {adadi2018peeking}
\bibfield{author}{\bibinfo{person}{Amina Adadi} {and} \bibinfo{person}{Mohammed
  Berrada}.} \bibinfo{year}{2018}\natexlab{}.
\newblock \showarticletitle{Peeking inside the black-box: A survey on
  Explainable Artificial Intelligence (XAI)}.
\newblock \bibinfo{journal}{\emph{IEEE Access}}  \bibinfo{volume}{6}
  (\bibinfo{year}{2018}), \bibinfo{pages}{52138--52160}.
\newblock


\bibitem[\protect\citeauthoryear{Bain and Muggleton}{Bain and
  Muggleton}{1994}]%
        {bain1994learning}
\bibfield{author}{\bibinfo{person}{Michael Bain} {and} \bibinfo{person}{Stephen
  Muggleton}.} \bibinfo{year}{1994}\natexlab{}.
\newblock \showarticletitle{Learning optimal chess strategies}. In
  \bibinfo{booktitle}{\emph{Machine intelligence 13}}. Oxford University Press,
  Inc., \bibinfo{pages}{291--309}.
\newblock


\bibitem[\protect\citeauthoryear{Bourgeois, Van Der~Linden, Kortuem, Price, and
  Rimmer}{Bourgeois et~al\mbox{.}}{2014}]%
        {bourgeois2014conversations}
\bibfield{author}{\bibinfo{person}{Jacky Bourgeois}, \bibinfo{person}{Janet Van
  Der~Linden}, \bibinfo{person}{Gerd Kortuem}, \bibinfo{person}{Blaine~A
  Price}, {and} \bibinfo{person}{Christopher Rimmer}.}
  \bibinfo{year}{2014}\natexlab{}.
\newblock \showarticletitle{Conversations with my washing machine: an
  in-the-wild study of demand shifting with self-generated energy}. In
  \bibinfo{booktitle}{\emph{Proceedings of the 2014 ACM International Joint
  Conference on Pervasive and Ubiquitous Computing}}. ACM,
  \bibinfo{pages}{459--470}.
\newblock


\bibitem[\protect\citeauthoryear{Boyer, Phillips, Ingram, Ha, Wallis, Vouk, and
  Lester}{Boyer et~al\mbox{.}}{2011}]%
        {boyer2011investigating}
\bibfield{author}{\bibinfo{person}{Kristy~Elizabeth Boyer},
  \bibinfo{person}{Robert Phillips}, \bibinfo{person}{Amy Ingram},
  \bibinfo{person}{Eun~Young Ha}, \bibinfo{person}{Michael Wallis},
  \bibinfo{person}{Mladen Vouk}, {and} \bibinfo{person}{James Lester}.}
  \bibinfo{year}{2011}\natexlab{}.
\newblock \showarticletitle{Investigating the relationship between dialogue
  structure and tutoring effectiveness: a hidden Markov modeling approach}.
\newblock \bibinfo{journal}{\emph{International Journal of Artificial
  Intelligence in Education}} \bibinfo{volume}{21}, \bibinfo{number}{1-2}
  (\bibinfo{year}{2011}), \bibinfo{pages}{65--81}.
\newblock


\bibitem[\protect\citeauthoryear{Caruana, Lou, Gehrke, Koch, Sturm, and
  Elhadad}{Caruana et~al\mbox{.}}{2015}]%
        {caruana2015intelligible}
\bibfield{author}{\bibinfo{person}{Rich Caruana}, \bibinfo{person}{Yin Lou},
  \bibinfo{person}{Johannes Gehrke}, \bibinfo{person}{Paul Koch},
  \bibinfo{person}{Marc Sturm}, {and} \bibinfo{person}{Noemie Elhadad}.}
  \bibinfo{year}{2015}\natexlab{}.
\newblock \showarticletitle{Intelligible models for healthcare: Predicting
  pneumonia risk and hospital 30-day readmission}. In
  \bibinfo{booktitle}{\emph{Proceedings of the 21th ACM SIGKDD International
  Conference on Knowledge Discovery and Data Mining}}. ACM,
  \bibinfo{pages}{1721--1730}.
\newblock


\bibitem[\protect\citeauthoryear{Charness, Tuffiash, Krampe, Reingold, and
  Vasyukova}{Charness et~al\mbox{.}}{2005}]%
        {charness2005role}
\bibfield{author}{\bibinfo{person}{Neil Charness}, \bibinfo{person}{Michael
  Tuffiash}, \bibinfo{person}{Ralf Krampe}, \bibinfo{person}{Eyal Reingold},
  {and} \bibinfo{person}{Ekaterina Vasyukova}.}
  \bibinfo{year}{2005}\natexlab{}.
\newblock \showarticletitle{The role of deliberate practice in chess
  expertise}.
\newblock \bibinfo{journal}{\emph{Applied Cognitive Psychology}}
  \bibinfo{volume}{19}, \bibinfo{number}{2} (\bibinfo{year}{2005}),
  \bibinfo{pages}{151--165}.
\newblock


\bibitem[\protect\citeauthoryear{Cheverst, Byun, Fitton, Sas, Kray, and
  Villar}{Cheverst et~al\mbox{.}}{2005}]%
        {cheverst2005exploring}
\bibfield{author}{\bibinfo{person}{Keith Cheverst}, \bibinfo{person}{Hee~Eon
  Byun}, \bibinfo{person}{Dan Fitton}, \bibinfo{person}{Corina Sas},
  \bibinfo{person}{Chris Kray}, {and} \bibinfo{person}{Nicolas Villar}.}
  \bibinfo{year}{2005}\natexlab{}.
\newblock \showarticletitle{Exploring issues of user model transparency and
  proactive behaviour in an office environment control system}.
\newblock \bibinfo{journal}{\emph{User Modeling and User-Adapted Interaction}}
  \bibinfo{volume}{15}, \bibinfo{number}{3-4} (\bibinfo{year}{2005}),
  \bibinfo{pages}{235--273}.
\newblock


\bibitem[\protect\citeauthoryear{Costanza, Fischer, Colley, Rodden, Ramchurn,
  and Jennings}{Costanza et~al\mbox{.}}{2014}]%
        {costanza2014doing}
\bibfield{author}{\bibinfo{person}{Enrico Costanza}, \bibinfo{person}{Joel~E
  Fischer}, \bibinfo{person}{James~A Colley}, \bibinfo{person}{Tom Rodden},
  \bibinfo{person}{Sarvapali~D Ramchurn}, {and} \bibinfo{person}{Nicholas~R
  Jennings}.} \bibinfo{year}{2014}\natexlab{}.
\newblock \showarticletitle{Doing the laundry with agents: a field trial of a
  future smart energy system in the home}. In
  \bibinfo{booktitle}{\emph{Proceedings of the SIGCHI Conference on Human
  Factors in Computing Systems}}. ACM, \bibinfo{pages}{813--822}.
\newblock


\bibitem[\protect\citeauthoryear{Dey}{Dey}{2018}]%
        {dey2018context}
\bibfield{author}{\bibinfo{person}{Anind~K Dey}.}
  \bibinfo{year}{2018}\natexlab{}.
\newblock \showarticletitle{Context-Aware Computing}.
\newblock In \bibinfo{booktitle}{\emph{Ubiquitous computing fundamentals}}.
  \bibinfo{publisher}{Chapman and Hall/CRC}, \bibinfo{pages}{335--366}.
\newblock


\bibitem[\protect\citeauthoryear{Ehsan, Tambwekar, Chan, Harrison, and
  Riedl}{Ehsan et~al\mbox{.}}{2019}]%
        {ehsan2019automated}
\bibfield{author}{\bibinfo{person}{Upol Ehsan}, \bibinfo{person}{Pradyumna
  Tambwekar}, \bibinfo{person}{Larry Chan}, \bibinfo{person}{Brent Harrison},
  {and} \bibinfo{person}{Mark~O Riedl}.} \bibinfo{year}{2019}\natexlab{}.
\newblock \showarticletitle{Automated rationale generation: a technique for
  explainable AI and its effects on human perceptions}. In
  \bibinfo{booktitle}{\emph{Proceedings of the 24th International Conference on
  Intelligent User Interfaces}}. ACM, \bibinfo{pages}{263--274}.
\newblock


\bibitem[\protect\citeauthoryear{Feng and Boyd-Graber}{Feng and
  Boyd-Graber}{2019}]%
        {feng2019can}
\bibfield{author}{\bibinfo{person}{Shi Feng} {and} \bibinfo{person}{Jordan
  Boyd-Graber}.} \bibinfo{year}{2019}\natexlab{}.
\newblock \showarticletitle{What can ai do for me?: evaluating machine learning
  interpretations in cooperative play}. In
  \bibinfo{booktitle}{\emph{Proceedings of the 24th International Conference on
  Intelligent User Interfaces}}. ACM, \bibinfo{pages}{229--239}.
\newblock


\bibitem[\protect\citeauthoryear{Fong and Vedaldi}{Fong and Vedaldi}{2017}]%
        {fong2017interpretable}
\bibfield{author}{\bibinfo{person}{Ruth~C Fong} {and} \bibinfo{person}{Andrea
  Vedaldi}.} \bibinfo{year}{2017}\natexlab{}.
\newblock \showarticletitle{Interpretable explanations of black boxes by
  meaningful perturbation}. In \bibinfo{booktitle}{\emph{Proceedings of the
  IEEE International Conference on Computer Vision}}.
  \bibinfo{pages}{3429--3437}.
\newblock


\bibitem[\protect\citeauthoryear{Fox, Long, and Magazzeni}{Fox
  et~al\mbox{.}}{2017}]%
        {fox2017explainable}
\bibfield{author}{\bibinfo{person}{Maria Fox}, \bibinfo{person}{Derek Long},
  {and} \bibinfo{person}{Daniele Magazzeni}.} \bibinfo{year}{2017}\natexlab{}.
\newblock \showarticletitle{Explainable planning}.
\newblock \bibinfo{journal}{\emph{arXiv preprint arXiv:1709.10256}}
  (\bibinfo{year}{2017}).
\newblock


\bibitem[\protect\citeauthoryear{Guidotti, Monreale, Ruggieri, Turini,
  Giannotti, and Pedreschi}{Guidotti et~al\mbox{.}}{2018}]%
        {10.1145/3236009}
\bibfield{author}{\bibinfo{person}{Riccardo Guidotti}, \bibinfo{person}{Anna
  Monreale}, \bibinfo{person}{Salvatore Ruggieri}, \bibinfo{person}{Franco
  Turini}, \bibinfo{person}{Fosca Giannotti}, {and} \bibinfo{person}{Dino
  Pedreschi}.} \bibinfo{year}{2018}\natexlab{}.
\newblock \showarticletitle{A Survey of Methods for Explaining Black Box
  Models}.
\newblock \bibinfo{journal}{\emph{ACM Comput. Surv.}} \bibinfo{volume}{51},
  \bibinfo{number}{5}, Article \bibinfo{articleno}{Article 93}
  (\bibinfo{date}{Aug.} \bibinfo{year}{2018}), \bibinfo{numpages}{42}~pages.
\newblock
\showISSN{0360-0300}
\urldef\tempurl%
\url{https://doi.org/10.1145/3236009}
\showDOI{\tempurl}


\bibitem[\protect\citeauthoryear{Gunning}{Gunning}{2017}]%
        {gunning2017explainable}
\bibfield{author}{\bibinfo{person}{David Gunning}.}
  \bibinfo{year}{2017}\natexlab{}.
\newblock \showarticletitle{Explainable artificial intelligence (xai)}.
\newblock \bibinfo{journal}{\emph{Defense Advanced Research Projects Agency
  (DARPA), nd Web}}  \bibinfo{volume}{2} (\bibinfo{year}{2017}).
\newblock


\bibitem[\protect\citeauthoryear{Gunning and Aha}{Gunning and Aha}{2019}]%
        {gunning2019darpa}
\bibfield{author}{\bibinfo{person}{David Gunning} {and}
  \bibinfo{person}{David~W Aha}.} \bibinfo{year}{2019}\natexlab{}.
\newblock \showarticletitle{DARPA's Explainable Artificial Intelligence
  Program}.
\newblock \bibinfo{journal}{\emph{AI Magazine}} \bibinfo{volume}{40},
  \bibinfo{number}{2} (\bibinfo{year}{2019}), \bibinfo{pages}{44--58}.
\newblock


\bibitem[\protect\citeauthoryear{Hilles and Naser}{Hilles and Naser}{2017}]%
        {hilles2017knowledge}
\bibfield{author}{\bibinfo{person}{Mohanad~M Hilles} {and}
  \bibinfo{person}{Samy S~Abu Naser}.} \bibinfo{year}{2017}\natexlab{}.
\newblock \showarticletitle{Knowledge-based Intelligent Tutoring System for
  Teaching Mongo Database}.
\newblock  (\bibinfo{year}{2017}).
\newblock


\bibitem[\protect\citeauthoryear{Hsu}{Hsu}{1999}]%
        {hsu1999ibm}
\bibfield{author}{\bibinfo{person}{Feng-hsiung Hsu}.}
  \bibinfo{year}{1999}\natexlab{}.
\newblock \showarticletitle{IBM's deep blue chess grandmaster chips}.
\newblock \bibinfo{journal}{\emph{IEEE Micro}} \bibinfo{volume}{19},
  \bibinfo{number}{2} (\bibinfo{year}{1999}), \bibinfo{pages}{70--81}.
\newblock


\bibitem[\protect\citeauthoryear{Kermany, Goldbaum, Cai, Valentim, Liang,
  Baxter, McKeown, Yang, Wu, Yan, et~al\mbox{.}}{Kermany et~al\mbox{.}}{2018}]%
        {kermany2018identifying}
\bibfield{author}{\bibinfo{person}{Daniel~S Kermany}, \bibinfo{person}{Michael
  Goldbaum}, \bibinfo{person}{Wenjia Cai}, \bibinfo{person}{Carolina~CS
  Valentim}, \bibinfo{person}{Huiying Liang}, \bibinfo{person}{Sally~L Baxter},
  \bibinfo{person}{Alex McKeown}, \bibinfo{person}{Ge Yang},
  \bibinfo{person}{Xiaokang Wu}, \bibinfo{person}{Fangbing Yan},
  {et~al\mbox{.}}} \bibinfo{year}{2018}\natexlab{}.
\newblock \showarticletitle{Identifying medical diagnoses and treatable
  diseases by image-based deep learning}.
\newblock \bibinfo{journal}{\emph{Cell}} \bibinfo{volume}{172},
  \bibinfo{number}{5} (\bibinfo{year}{2018}), \bibinfo{pages}{1122--1131}.
\newblock


\bibitem[\protect\citeauthoryear{Koh and Liang}{Koh and Liang}{2017}]%
        {10.5555/3305381.3305576}
\bibfield{author}{\bibinfo{person}{Pang~Wei Koh} {and} \bibinfo{person}{Percy
  Liang}.} \bibinfo{year}{2017}\natexlab{}.
\newblock \showarticletitle{Understanding Black-Box Predictions via Influence
  Functions}. In \bibinfo{booktitle}{\emph{Proceedings of the 34th
  International Conference on Machine Learning - Volume 70}}
  \emph{(\bibinfo{series}{ICML’17})}. \bibinfo{publisher}{JMLR.org},
  \bibinfo{pages}{1885–1894}.
\newblock


\bibitem[\protect\citeauthoryear{Lakkaraju, Bach, and Leskovec}{Lakkaraju
  et~al\mbox{.}}{2016}]%
        {lakkaraju2016interpretable}
\bibfield{author}{\bibinfo{person}{Himabindu Lakkaraju},
  \bibinfo{person}{Stephen~H Bach}, {and} \bibinfo{person}{Jure Leskovec}.}
  \bibinfo{year}{2016}\natexlab{}.
\newblock \showarticletitle{Interpretable decision sets: A joint framework for
  description and prediction}. In \bibinfo{booktitle}{\emph{Proceedings of the
  22nd ACM SIGKDD international conference on knowledge discovery and data
  mining}}. ACM, \bibinfo{pages}{1675--1684}.
\newblock


\bibitem[\protect\citeauthoryear{Letham, Rudin, McCormick, Madigan,
  et~al\mbox{.}}{Letham et~al\mbox{.}}{2015}]%
        {letham2015interpretable}
\bibfield{author}{\bibinfo{person}{Benjamin Letham}, \bibinfo{person}{Cynthia
  Rudin}, \bibinfo{person}{Tyler~H McCormick}, \bibinfo{person}{David Madigan},
  {et~al\mbox{.}}} \bibinfo{year}{2015}\natexlab{}.
\newblock \showarticletitle{Interpretable classifiers using rules and bayesian
  analysis: Building a better stroke prediction model}.
\newblock \bibinfo{journal}{\emph{The Annals of Applied Statistics}}
  \bibinfo{volume}{9}, \bibinfo{number}{3} (\bibinfo{year}{2015}),
  \bibinfo{pages}{1350--1371}.
\newblock


\bibitem[\protect\citeauthoryear{Mahdi, Alhabbash, and Naser}{Mahdi
  et~al\mbox{.}}{2016}]%
        {mahdi2016intelligent}
\bibfield{author}{\bibinfo{person}{Ali~O Mahdi}, \bibinfo{person}{Mohammed~I
  Alhabbash}, {and} \bibinfo{person}{Samy S~Abu Naser}.}
  \bibinfo{year}{2016}\natexlab{}.
\newblock \showarticletitle{An intelligent tutoring system for teaching
  advanced topics in information security}.
\newblock  (\bibinfo{year}{2016}).
\newblock


\bibitem[\protect\citeauthoryear{Montfort and Bogost}{Montfort and
  Bogost}{2009}]%
        {montfort2009racing}
\bibfield{author}{\bibinfo{person}{Nick Montfort} {and} \bibinfo{person}{Ian
  Bogost}.} \bibinfo{year}{2009}\natexlab{}.
\newblock \bibinfo{booktitle}{\emph{Racing the beam: The Atari video computer
  system}}.
\newblock \bibinfo{publisher}{Mit Press}.
\newblock


\bibitem[\protect\citeauthoryear{Ribeiro, Singh, and Guestrin}{Ribeiro
  et~al\mbox{.}}{2016}]%
        {ribeiro2016should}
\bibfield{author}{\bibinfo{person}{Marco~Tulio Ribeiro},
  \bibinfo{person}{Sameer Singh}, {and} \bibinfo{person}{Carlos Guestrin}.}
  \bibinfo{year}{2016}\natexlab{}.
\newblock \showarticletitle{Why should i trust you?: Explaining the predictions
  of any classifier}. In \bibinfo{booktitle}{\emph{Proceedings of the 22nd ACM
  SIGKDD international conference on knowledge discovery and data mining}}.
  ACM, \bibinfo{pages}{1135--1144}.
\newblock


\bibitem[\protect\citeauthoryear{Simon and Chase}{Simon and Chase}{1988}]%
        {simon1988skill}
\bibfield{author}{\bibinfo{person}{Herbert Simon} {and}
  \bibinfo{person}{William Chase}.} \bibinfo{year}{1988}\natexlab{}.
\newblock \showarticletitle{Skill in chess}.
\newblock In \bibinfo{booktitle}{\emph{Computer chess compendium}}.
  \bibinfo{publisher}{Springer}, \bibinfo{pages}{175--188}.
\newblock


\bibitem[\protect\citeauthoryear{Stockfish}{Stockfish}{[n.d.]}]%
        {Stockfish}
\bibfield{author}{\bibinfo{person}{Stockfish}.}
  \bibinfo{year}{[n.d.]}\natexlab{}.
\newblock
\newblock
\newblock
\shownote{\url{http://stockfishchess.org/}.}


\bibitem[\protect\citeauthoryear{Sutcliffe, van Assche, and Benyon}{Sutcliffe
  et~al\mbox{.}}{2016}]%
        {sutcliffe2016domain}
\bibfield{author}{\bibinfo{person}{Alistair~G Sutcliffe},
  \bibinfo{person}{Frans van Assche}, {and} \bibinfo{person}{David Benyon}.}
  \bibinfo{year}{2016}\natexlab{}.
\newblock \bibinfo{booktitle}{\emph{Domain knowledge for interactive system
  design}}.
\newblock \bibinfo{publisher}{Springer}.
\newblock


\bibitem[\protect\citeauthoryear{Teichmann, Weber, Zoellner, Cipolla, and
  Urtasun}{Teichmann et~al\mbox{.}}{2018}]%
        {teichmann2018multinet}
\bibfield{author}{\bibinfo{person}{Marvin Teichmann}, \bibinfo{person}{Michael
  Weber}, \bibinfo{person}{Marius Zoellner}, \bibinfo{person}{Roberto Cipolla},
  {and} \bibinfo{person}{Raquel Urtasun}.} \bibinfo{year}{2018}\natexlab{}.
\newblock \showarticletitle{Multinet: Real-time joint semantic reasoning for
  autonomous driving}. In \bibinfo{booktitle}{\emph{2018 IEEE Intelligent
  Vehicles Symposium (IV)}}. IEEE, \bibinfo{pages}{1013--1020}.
\newblock


\bibitem[\protect\citeauthoryear{Wu, Hughes, Parbhoo, Zazzi, Roth, and
  Doshi-Velez}{Wu et~al\mbox{.}}{2018}]%
        {wu2018beyond}
\bibfield{author}{\bibinfo{person}{Mike Wu}, \bibinfo{person}{Michael~C
  Hughes}, \bibinfo{person}{Sonali Parbhoo}, \bibinfo{person}{Maurizio Zazzi},
  \bibinfo{person}{Volker Roth}, {and} \bibinfo{person}{Finale Doshi-Velez}.}
  \bibinfo{year}{2018}\natexlab{}.
\newblock \showarticletitle{Beyond sparsity: Tree regularization of deep models
  for interpretability}. In \bibinfo{booktitle}{\emph{Thirty-Second AAAI
  Conference on Artificial Intelligence}}.
\newblock


\bibitem[\protect\citeauthoryear{Zhang, Cao, Shi, Wu, and Zhu}{Zhang
  et~al\mbox{.}}{2018}]%
        {zhang2018interpreting}
\bibfield{author}{\bibinfo{person}{Quanshi Zhang}, \bibinfo{person}{Ruiming
  Cao}, \bibinfo{person}{Feng Shi}, \bibinfo{person}{Ying~Nian Wu}, {and}
  \bibinfo{person}{Song-Chun Zhu}.} \bibinfo{year}{2018}\natexlab{}.
\newblock \showarticletitle{Interpreting cnn knowledge via an explanatory
  graph}. In \bibinfo{booktitle}{\emph{Thirty-Second AAAI Conference on
  Artificial Intelligence}}.
\newblock


\bibitem[\protect\citeauthoryear{Zhang, Yang, Ma, and Wu}{Zhang
  et~al\mbox{.}}{2019}]%
        {zhang2019interpreting}
\bibfield{author}{\bibinfo{person}{Quanshi Zhang}, \bibinfo{person}{Yu Yang},
  \bibinfo{person}{Haotian Ma}, {and} \bibinfo{person}{Ying~Nian Wu}.}
  \bibinfo{year}{2019}\natexlab{}.
\newblock \showarticletitle{Interpreting cnns via decision trees}. In
  \bibinfo{booktitle}{\emph{Proceedings of the IEEE Conference on Computer
  Vision and Pattern Recognition}}. \bibinfo{pages}{6261--6270}.
\newblock


\bibitem[\protect\citeauthoryear{Zhang and Zhu}{Zhang and Zhu}{2018}]%
        {zhang2018visual}
\bibfield{author}{\bibinfo{person}{Quan-shi Zhang} {and}
  \bibinfo{person}{Song-Chun Zhu}.} \bibinfo{year}{2018}\natexlab{}.
\newblock \showarticletitle{Visual interpretability for deep learning: a
  survey}.
\newblock \bibinfo{journal}{\emph{Frontiers of Information Technology \&
  Electronic Engineering}} \bibinfo{volume}{19}, \bibinfo{number}{1}
  (\bibinfo{year}{2018}), \bibinfo{pages}{27--39}.
\newblock


\end{thebibliography}

\appendix

\end{document}